\documentclass[twocolumn,showpacs,preprintnumbers,amsmath,amssymb]{revtex4}
\usepackage{graphicx}
\usepackage{dcolumn} 
\usepackage{bm}      
\usepackage{epsf,psfig}

\begin{document}








{\bf Reply to Millis {\it et al.}:}
In a recent paper \cite{us} we showed the equivalence
of two seemingly contradictory theories \cite{us2000,millis2002} 
on Griffiths-McCoy singularities (GMS) in metallic antiferromagnets close to 
a quantum critical point (QCP). Two generic features of these theories are 
the presence of ordered magnetic droplets in the paramagnetic
phase, and a non-universal crossover from GMS to superparamagnetism 
below a temperature $T^* = \omega_0 \exp\{-C_2\}$, 
where $\omega_0$ is a cut-off energy and $C_2$ is a non-universal constant 
dependent on the electronic damping of the spin excitations.
In a recent comment, Millis {\it et al.}
\cite{millisetal} argue that in heavy-fermion materials the electronic
damping is large 
leading to the freezing of locally magnetically ordered 
droplets at high temperatures. In this reply we show that
this 
conclusion is based on a treatment of the problem
of disorder close to a QCP which is not self-consistent. 
We argue that a self-consistent treatment of the ordered droplets
must lead to weak damping and to a large region of GMS behavior,
in agreement with the results of ref.~\cite{us2000}.

Clean heavy-fermion materials can be described by the Kondo
lattice model:
$
H = H_0 + \sum_{i} J_{K} {\bf S}_i \cdot {\bf s}_i
$, where $H_0 = \sum_{{\bf k},\sigma} \epsilon_{{\bf k}} c^{\dag}_{{\bf
    k},\sigma} c_{{\bf k},\sigma}$ is the conduction electron Hamiltonian 
($c^{\dag}_{{\bf k},\sigma}$ and $c_{{\bf k},\sigma}$ are the creation
and annihilation operators for electrons with momentum ${\bf k}$, 
spin $\sigma$, and energy
$\epsilon_{\bf k}$), $J_K$ is the Kondo coupling at site ${\bf R}_i$ 
between localized spins ${\bf S}_i$, and conduction electron spin 
$s^a_i = (\hbar/2) \sum_{\alpha,\beta}  c^{\dag}_{i,\alpha}
    \sigma^a_{\alpha,\beta} c_{i,\beta}$ ($\sigma^a_{\alpha,\beta}$
with $a=x,y,z$ are Pauli matrices). 
In the absence of a full theory of the heavy-fermion state close
to a QCP we follow the argument proposed by Doniach \cite{doniach}
and recent results from extended dynamical mean-field theory (EDMFT) 
\cite{sun,si}. The Kondo coupling is responsible
for two competing effects that originate a QCP: 
the RKKY interaction, responsible for magnetic ordering, 
with characteristic energy scale $T_c \sim g^2 E_F$,
and the Kondo effect, responsible for the heavy-fermion paramagnetic phase 
below a temperature $T_K \sim E_F \exp\{-1/g\}$, where 
$E_F$ is the Fermi energy.
The dimensionless coupling $g \sim J_K N(0)$, where  
$N(0)$ is the electronic density of states, 
determines the phase diagram at $T=0$: when $g<g_c$ the system
is magnetically ordered and the Kondo effect is suppressed;
when $g>g_c$ Kondo singlets are formed and the system
is paramagnetic. EDMFT calculations 
\cite{sun,si} find that $g_c$ is order of unit ($\sim {\cal O}(1)$). 
Inside the magnetic ordered phase, $g < g_c$,
the system is an ordinary Fermi liquid with $N_{g<g_c}(0) \propto 1/E_F$ \cite{si}.
In the paramagnetic phase, $g > g_c$, the Kondo effect leads 
to a large renormalization of the density of states, $N_{g>g_c}(0) \propto 1/T_K$,
and to a large effective mass, $m^*$: $N_{g>g_c}(0) \propto m^*_{g>g_c} 
\gg N_{g<g_c}(0) \propto m^*_{g<g_c}$, since $E_F \gg T_K$.

As we have shown in ref.~\cite{us2000}, {\it in the presence of disorder}
(either by modification of the electronic degrees of freedom via $N(0)$ or by
the change in the magnetic interactions via $J_K$) the dimensionless
coupling becomes position dependent, $g(i)$, and is therefore 
statistically distributed.  Even if the material is in the
paramagnetic phase 
there will be regions where locally $g(i)<g_c$ leading to the formation of 
magnetically ordered droplets of size $N$. 
It was also shown \cite{us2000} that for antiferromagnetic droplets there
is a critical droplet size $N_c \propto 1/g^2$ such that for $N>N_c$
the droplets freeze at low temperatures due to the electronic damping 
\cite{note}. 

Ref.~\cite{millis2002} 
finds that the density of states within the
droplet is given by $N_{g>g_c}(0)$ so that $g(i) \approx J_K/T_K \sim {\cal O}(1)$,
implying that $N_c \sim {\cal O}(1)$ and therefore concluding 
that essentially most droplets are frozen. This argument, however,
is incorrect. Firstly, to have an ordered droplet in first place 
one has to require that locally $g(i)<g_c \sim {\cal O}(1)$. Since
ref.~\cite{millisetal} assume that $g(i) \sim {\cal O}(1)$ the formation
of droplets is suppressed, in other words, if the local
coupling is too large, leading to large dissipation as argued in 
ref.~\cite{millisetal}, the RKKY interaction
is not able to stabilize local order and droplet formation. 
Secondly, and even more importantly, 
a droplet is a piece of the magnetically ordered phase 
and therefore the damping should be given by the {\it local}
density of states, $N_{g(i)<g_c}(0)$, which is not enhanced. 
In this case, as argued in ref.~\cite{us2000}, one has to use $N_{g<g_c}(0)$
for the droplet damping. Hence,
$g(i) \approx J_K/E_F \ll 1 < g_c \sim {\cal O}(1)$, 
consistent with $N_c \gg 1$ and with the existence of GMS. 

One can track down the inconsistency in the arguments of 
refs.~\cite{millis2002,millisetal} to the inappropriate use of
Hertz theory \cite{hertz}.
In Hertz approach one works with the order parameter alone
and the effects of dissipation originate on the trace, in the
partition function, of the electronic degrees of freedom.
The trace is carried out {\it perturbatively} 
in $g$ and leads to Landau damping with a damping rate, 
$\Gamma(\omega_n) \propto g^2 |\omega_n|$. Hence, 
it is questionable that this approach can be used to
describe situations where $g \sim {\cal O}(1)$, and therefore,
outside the applicability of perturbation theory.
In the presence of droplets the trace over electrons 
has do be done self-consistently
because the electronic properties are directly 
affected by the presence of local order in the system.
In ref.~\cite{millis2002} it was found that the local changes
in the electronic properties are not significant.
In that theory disorder enters only
through the order parameter and in the generation of the droplets. 
Thus, there is no reason to believe
that the dissipation rate is tied to a single energy scale $E_0 \sim T_K$
as proposed in ref.~\cite{millisetal}. 
In ref.~\cite{millis2002} 
the dissipation rate should be seen as free parameter of the theory and $T^*$
can be obtained from the experimental data. We should stress 
that  the microscopic theory of ref.~\cite{us2000} estimates that $N_c \approx 
10^5$ for systems like UCu$_{4-x}$Pd$_x$. This result supports the conclusion
that GMS can be the source of non-Fermi liquid behavior observed in a large
class of U and Ce intermetallics \cite{greg}.

\noindent
A.~H.~Castro Neto\\
{\small Department of Physics, Boston University, Boston, MA 02215}

\noindent
B.~A.~Jones\\
{\small IBM Research Center, San Jose, CA 95120}


\begin{thebibliography}{50}
\bibitem{note}Notice that in the insulating case $g \to 0$ and 
$N_c \to \infty$.
\end{thebibliography}
\end{document}